\documentclass[a4paper,11pt]{article}
\usepackage[utf8]{inputenc}
\usepackage[left=2.54cm,right=2.54cm,top=2.54cm,bottom=2.54cm]{geometry}

\usepackage{graphicx} 
\usepackage{subcaption}
\usepackage{float}
\usepackage{stackrel}
\usepackage{longtable}
\usepackage{multicol}
\usepackage{tabularx}
\usepackage{pdflscape}
\usepackage{array}
\usepackage{adjustbox}
\usepackage{amsmath,mathtools}
\usepackage[numbers]{natbib}
\usepackage[colorlinks=true,linkcolor=blue,urlcolor=blue,citecolor=blue]{hyperref} 
\providecommand{\keywords}[1]{\small\textbf{\textit{Keywords:}} #1}
\usepackage{authblk}

\title{Regional Price Dynamics and Market Integration in the U.S. Beef Industry: An Econometric Analysis}

\author[1,2,3,4]{Leonardo Manríquez-Méndez \thanks{\href{mailto:lmanriquez@ucsc.cl}{lmanriquez@ucsc.cl}}}

\affil[1]{Universidad Católica de la Santísima Concepción, Concepción, Chile}
\affil[2]{Universidad de Concepción, Concepción, Chile}
\affil[3]{Universidad del Bío-Bío, Concepción, Chile}
\affil[4]{Universidad Técnica Federico Santa María, Concepción, Chile}
\begin{document}

\maketitle

\begin{abstract}
The United States, a leading global producer and consumer of beef, continues to face substantial challenges in achieving price harmonization across its regional markets. This paper evaluates the validity of the Law of One Price (LOP) in the U.S. beef industry and investigates causal relationships among regional price dynamics. Through a series of econometric tests, we establish that regional price series are integrated of order one, displaying non-stationarity in levels and stationarity in first differences. The analysis reveals partial LOP compliance in the Northeast and West, while full convergence remains elusive at the national level. Although no region demonstrates persistent price leadership, Southern prices appear particularly sensitive to exogenous shocks. These findings reflect asymmetrical integration across U.S. beef markets and suggest the presence of structural frictions that hinder complete market unification.
\end{abstract}

\newpage

\section{Introduction}

The United States (U.S.) is the world's largest beef producer, and its market is primarily concentrated in two sectors: cow-calf operations and livestock feed. Beef cattle are bred in all U.S. states, and as of January 1, 2020, the cattle population totaled 94.4 million. The three states with the highest concentration of cattle are Texas (approximately 14 percent of the national total), Nebraska (approximately seven percent of the national total), and Kansas (approximately six percent of the national total).

American consumers are among the largest beef consumers globally, consuming approximately 57.2 pounds per capita in 2018 \cite{ERS2019}. However, projections from the Food and Agriculture Organization (FAO) indicate that while beef consumption in the U.S. will increase, it will grow at a slower rate than in developing countries. Specifically, beef consumption is expected to rise by approximately six percent in developed economies like the U.S. by 2026, compared to approximately 17 percent in developing economies. The overall trend in beef consumption is downward, primarily due to its high price compared to other meats, such as poultry.

Meat production is strongly driven by consumption, which is critical for farmers, meat processors, and other stakeholders. As price is a decisive factor in trade, understanding the dynamics of beef prices and the level of market integration in the U.S. is crucial. According to \cite{farm}, market integration significantly affects economic growth, alters the location of economic activity, and impacts the viability of small and large agricultural enterprises. This issue is of great concern to policymakers. Lack of market integration may indicate logistical inefficiencies, competitive issues within the retail supply chain, or regional transaction costs that ultimately distort consumer prices. Therefore, examining market integration is vital to ensure fair pricing, promote competition, and enhance overall market efficiency \cite{goletti1994maize}. However, \cite{Vollrath} notes that efforts to expand the agricultural market in the U.S. have sparked significant controversy within farming communities.

This paper aims to determine the existence of the Law of One Price (LOP) across regions in the U.S. beef market. As a secondary objective, it seeks to characterize U.S. regions as either influential or influenced based on their causal relationships with prices in other regions. To achieve these objectives, a multivariate analysis of beef price series in different U.S. regions will be conducted, specifically using a cointegration analysis through the Vector Error Correction Model (VECM).

The LOP, a price equilibrium condition, is fundamental to many international trade models. This law posits that commodity arbitrage ensures every good has a unique price, defined in a common global monetary unit. International trade models often assume the existence of the LOP. For instance, \cite{Baffes} examines the LOP in seven commodities across four countries, finding that the failure of the LOP as a long-term relationship is often specific to the price and time analyzed rather than a general issue. He also identifies transport costs as a primary factor in LOP failure. These findings align with studies by \cite{Stoll} and \cite{Goodwin}. Consequently, researchers have assessed market integration in various commodities. For example, \cite{weiner1991world}, \cite{gulen1997regionalization}, and \cite{alsaudi} tested the integration of oil markets.

In the agricultural sector, \cite{Kuiper} investigates six regional maize markets in Benin, finding a long-term relationship among them. \cite{Asche} explores the LOP in the global salmon market through cointegration analysis, concluding that the LOP applies to an international market with five salmon species. \cite{Vinuya} confirms the LOP in the shrimp market across Japan, the U.S., and the European Union, demonstrating strong market stability and evidence of the LOP. \cite{Katrakilidis}, using multivariate cointegration analysis and VECM, examines long-term relationships in milk markets across European countries, identifying causal effects and dominant markets that influence prices in submissive markets. Similarly, \cite{roman2020milk} finds evidence of the LOP in the EU milk market. \cite{clark2015law} supports the LOP in the Czech cereals market when accounting for structural breaks. \cite{kustiari2017market} finds evidence supporting the LOP in Indonesia's chili market.

In the meat market, \cite{Vollrath} examines border integration in U.S. and Canadian beef markets, using the LOP to verify market integration between both economies. \cite{Rumankova} investigates the inter-regional LOP in the Czech wholesale pork and beef markets, finding \textit{partial compliance} with the LOP, primarily due to transaction costs. Like \cite{Katrakilidis}, the author identifies dominant and submissive regions in these markets. This analysis aligns with \cite{Goodwin2}, who notes that wholesale meat markets adjust to shocks more than retail markets. \cite{Eskandarpour} analyzes the LOP in Iraq's meat market but finds insufficient evidence to support the hypothesis, largely due to inefficiencies and uncompetitive practices caused by varying tariffs on meat imports, which hinder domestic producers' ability to compete globally.

This article is structured as follows. Section 2 presents a conceptual framework for modeling the market integration study. Section 3 presents the data and main results. Finally, Section 4 discusses the main findings and implications for the U.S. beef market.

\section{Theoretical Framework}

The Law of One Price (LOP) postulates that international arbitrage in efficient markets implies that, for a homogeneous good, assuming the absence of transportation costs, prices (in the same currency) of $N$ markets are related according to:

\begin{equation}
p_{1t} = \beta_{2}p_{2t} + \beta_{3}p_{3t} + \dots + \beta_{N}p_{Nt},
\label{eq:lop1}
\end{equation}

where $\sum_{i=2}^{N}\beta_{i} = 1$. Equation \eqref{eq:lop1} represents the LOP in the strict sense, implying perfect market integration. This is understood as the existence of a single market in which prices are determined simultaneously, with differences due solely to transaction costs.

According to \cite{Sexton91}, breaches of the LOP primarily result from: the absence of price arbitrage; imperfect arbitrage; or imperfect competition. These factors explain varying degrees of market integration. If the degree of integration is \textit{weak}, deviations from equilibrium are likely permanent. Conversely, if the degree of integration is \textit{high}, it indicates that markets are efficient both internally and interrelationally. This efficiency allows prices to respond to variations in some markets, and in the long term, deviations from equilibrium are transitory, reflecting perfect long-term arbitrage. The LOP, as a long-term equilibrium relationship, takes the following form:

\begin{equation}
\beta_{1}p_{1t} + \beta_{2}p_{2t} + \beta_{3}p_{3t} + \dots + \beta_{N}p_{Nt} = \sum_{i=1}^{N}\beta_{i}p_{it} = 0,
\label{eq:LOP_lp}
\end{equation}

where $\sum_{i=1}^{N}\beta_{i} = 0$. In \eqref{eq:LOP_lp}, short-term deviations from the equilibrium in \eqref{eq:lop1} are permitted but are transient. When $\sum_{i=1}^{N}\beta_{i}p_{it}$ is a stationary process, it indicates a long-term equilibrium.

This approach is valuable because it allows the use of cointegration theory to verify LOP compliance. The most recognized methods are those proposed by \cite{EngleGranger87} and \cite{Johansen88}. The primary difference is that the former assumes a single cointegration relationship among the $N$ series and imposes endogeneity on one series, while the latter, more general approach, allows for up to $N-1$ cointegration vectors, with all $N$ series potentially endogenous and explanatory simultaneously. This paper adopts the Johansen approach to test for long-run price relationships consistent with the LOP in the U.S. beef market.

\subsection{Econometric Specification}

Given that cointegration analysis applies to nonstationary series, the first step is to confirm the nonstationary nature of the price series. This paper employs the Augmented Dickey-Fuller (ADF) and Phillips-Perron (PP) tests. The ADF test is specified as:

\begin{equation}
\Delta p_{it} = \alpha + \pi p_{i,t-1} + \delta t + \sum_{\gamma=1}^{k} \psi \Delta p_{i,t-\gamma} + \epsilon_{t},
\label{eq:adf}
\end{equation}

where $k$ represents the number of lags to include, and the constant and/or trend terms may be omitted depending on the series' nature. The PP test is specified as:

\begin{equation}
\Delta p_{t} = \alpha + \gamma t + \pi p_{t} + \epsilon_{t},
\end{equation}

where $\epsilon_{t} \sim I(0)$ and may exhibit autocorrelation and heteroscedasticity. For both ADF and PP tests, the null hypothesis is nonstationarity, i.e., $H_{0}: \pi = 0$. These tests are applied to price series at levels and in first differences.

Once the price series are confirmed to be integrated of order one, the existence of cointegration relationships is evaluated using the Johansen method.\footnote{This test is sensitive to: (1) the presence of a trend in the original series, (2) the inclusion of an intercept in the cointegration vector, (3) the presence of a trend in the cointegration vector, and (4) the inclusion of an intercept in the VAR model. The specific variant of the test used in this study will be specified as needed.} The Johansen method is based on the Vector Autoregressive (VAR) model in first differences:

\begin{equation}
\Delta p_{t} = \Pi p_{t-1} + \sum_{i=1}^{k-1} \Gamma_{i} \Delta p_{t-i} + \Phi D_{t} + \epsilon_{t},
\label{eq:var_1dif}
\end{equation}

where $\Pi = -(I - \sum_{i=1}^{k} \pi_{i})$, $\Gamma_{i} = -\sum_{j=i+1}^{k} \pi_{j}$, and $D$ is a vector of deterministic variables. If the matrix $\Pi$ in \eqref{eq:var_1dif} has rank $r$, with $0 < r < N$, it can be decomposed as $\Pi = \alpha \beta'$, where $\alpha$ and $\beta$ are $N \times r$ matrices. The elements of $\alpha$ represent adjustment speeds, and the elements of $\beta$ are the coefficients of the cointegration vectors. If the rank of $\Pi$ satisfies $0 < r < N$, a Vector Error Correction Model (VECM) with $r$ cointegration relationships is specified as:

\begin{equation}
\Delta p_{t} = \alpha \beta' p_{t-1} + \sum_{i=1}^{k-1} \Gamma_{i} \Delta p_{t-i} + \Phi D_{t} + \epsilon_{t},
\label{eq:vec}
\end{equation}

After confirming that the price series are integrated of order one, determining the optimal number of lags in the VAR model, and verifying at the univariate and multivariate levels that the residuals are not autocorrelated and are normally distributed, Johansen’s methodology is applied to determine the number of cointegration vectors, $r$.

The number of cointegration vectors, $r$, is determined using the methodology of \cite{Johansen95}, with the maximum eigenvalue and trace criteria. Inference on the vectors is conducted following \cite{Juselius}.

The primary advantage of Johansen’s cointegration approach for testing market integration and the LOP is its ability to test hypotheses on the coefficients of the $\alpha$ and $\beta$ matrices using likelihood ratio tests. To test the LOP, restrictions are imposed on the parameters of $\beta$. In a bivariate system of two price series, the LOP is tested by imposing the restriction $\beta' = [1, -1]$. Since $\beta$ contains long-term parameters, this constitutes a statistical test of the LOP’s validity. In the multivariate case, assuming the rank of $\Pi$ is $N-1$, the LOP is tested by ensuring the columns of $\beta$ sum to zero. For four price series, the matrix is represented as:

\begin{equation}
\label{eq:ho_lop}
\beta' =
\begin{bmatrix}
1 & -\beta_{1} & 0 & 0 \\
1 & 0 & -\beta_{2} & 0 \\
1 & 0 & 0 & -\beta_{3} \\
\end{bmatrix}.
\end{equation}

Each row represents a cointegration relationship. The adjustment parameters in $\alpha$ are also of interest, as they relate to weak exogeneity. If all adjustment parameters for a variable are zero, that variable is weakly exogenous to the long-term parameters in the remaining equations, implying that other price variables do not influence it in the long term.

The ability to identify weak exogeneity in certain price series enables the detection of potential market and price leadership based on the causal direction provided by the adjustment parameters.

\section{Data and Results}

In this paper, we analyze the logarithms of monthly beef prices \footnote{We consider the category of uncooked beef fillet, per pound (453.6 grams). We acknowledge that BLS reports listed prices rather than transaction-level data, which may introduce volatility and noise in prices.} published by the Bureau of Labor Statistics (BLS) for different regions of the U.S. from January 1998 to November 2020. The regions are defined according to BLS classifications as follows:

\begin{itemize}
\item Northeast ($\text{log}p_{t}^{NE}$): Connecticut, Maine, Massachusetts, New Hampshire, New Jersey, New York, Pennsylvania, Rhode Island, and Vermont.
\item South ($\text{log}p_{t}^{SO}$): Alabama, Arkansas, Delaware, the District of Columbia, Florida, Georgia, Kentucky, Louisiana, Maryland, Mississippi, North Carolina, Oklahoma, South Carolina, Tennessee, Texas, Virginia, and West Virginia.
\item Midwest ($\text{log}p_{t}^{MW}$): Illinois, Indiana, Iowa, Kansas, Michigan, Minnesota, Missouri, Nebraska, North Dakota, Ohio, South Dakota, and Wisconsin.
\item West ($\text{log}p_{t}^{WE}$): Alaska, Arizona, California, Colorado, Hawaii, Idaho, Montana, Nevada, New Mexico, Oregon, Utah, Washington, and Wyoming.
\end{itemize}

\begin{figure}[H][H]
\centering
\caption{Monthly Retail Fillet Beef Prices (Log Scale), 1998–2020, BLS}
\includegraphics{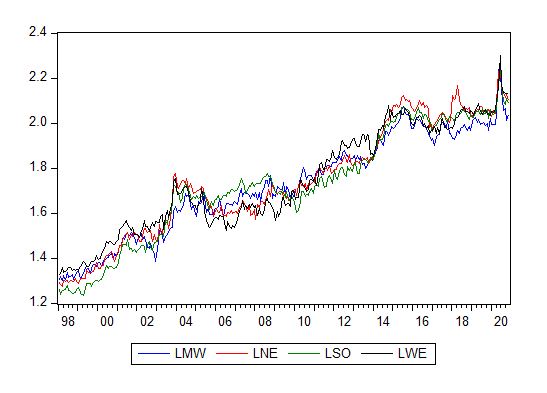}
\label{fig:comportamiento_prices}
\subcaption*{Note: The analysis period spans January 1998 to November 2020. The vertical axis shows the logarithm of the series, and the horizontal axis represents time.}
\end{figure}

Figure \ref{fig:comportamiento_prices} shows the behavior of the logarithm of each price series.

\subsection{Unit Root Test}

Table \ref{tab:df_pp} shows that the price series in the four U.S. regions are nonstationary in levels but stationary in first differences, indicating that the price logarithms are integrated of order one.

\begin{table}[H]
\centering
\caption{Unit Root Test}
\begin{tabular}{ccccc}\hline\hline
Variable (log) & ADF (level) & PP (level) & ADF (first diff.) & PP (first diff.)\\ \hline
$p^{MW}$ & -1.02 (0.74) & -1.21 (0.66) & -8.05 (0.00) & -22.98 (0.00) \\
$p^{NE}$ & -1.19 (0.67) & -1.12 (0.70) & -17.49 (0.00) & -17.79 (0.00) \\
$p^{SO}$ & -1.25 (0.65) & -1.16 (0.69) & -11.77 (0.00) & -15.86 (0.00) \\
$p^{WE}$ & -0.84 (0.80) & -0.84 (0.80) & -11.82 (0.00) & -19.25 (0.00) \\ \hline\hline
\end{tabular}
\subcaption*{Note: The test includes an intercept. Critical values are -3.45 at 1\% and -2.87 at 5\% with an intercept \cite{MacKinnon1996}. P-values are in parentheses. The procedure is performed with a maximum lag of 15, selected by the AIC criterion.}
\label{tab:df_pp}
\end{table}

\subsection{VAR Model Estimation}

The VAR model is developed using the four logarithms of regional beef prices in the U.S., a constant, and a linear trend. Table \ref{tab:criterios_lagvar} presents various criteria for selecting the optimal order of the VAR model. Based on the AIC criterion, the model order is set to three.

\begin{table}[H]
\centering
\caption{Order Determination for VAR Model}
\begin{tabular}{lrrrrrr} \hline \hline
Lag & LogL & LR & FPE & AIC & SC & HQ \\ \hline
0 & 1797.025 & NA & 2.39e-11 & -13.10389 & -12.73072 & -12.95404 \\
1 & 2636.481 & 1610.512 & 5.37e-14 & -19.20356 & -18.61715 & -18.96808 \\
2 & 2688.852 & 98.92398 & 4.10e-14 & -19.47298 & -18.67333* & -19.15188* \\
3 & 2706.535 & 32.87676 & 4.06e-14* & -19.48544* & -18.47255 & -19.07871 \\
4 & 2721.632 & 27.62247* & 4.09e-14 & -19.47876 & -18.25263 & -18.98640 \\
5 & 2733.903 & 22.08676 & 4.20e-14 & -19.45113 & -18.01176 & -18.87314 \\ \hline\hline
\end{tabular}
\subcaption*{Note: * indicates the lag selected by the corresponding criterion.}
\label{tab:criterios_lagvar}
\end{table}

The initial VAR model is estimated as:

\begin{equation}
(I_{p} - \pi_{1}L - \pi_{2}L^{2} - \pi_{3}L^{3})p_{t} = c + \delta t + \mu_{t},
\label{eq:VAR1}
\end{equation}

where $p_{t} = [\text{log}p_{t}^{MW}, \text{log}p_{t}^{NE}, \text{log}p_{t}^{SO}, \text{log}p_{t}^{WE}]'$, $t$ represents a linear trend, and the $\pi_{i}$ matrices are $4 \times 4$. The estimation results are presented in Table \ref{tab:VAR_original}. As shown in Tables \ref{tab:serial_VAR3} and \ref{tab:week1}, the model exhibits error autocorrelation and nonnormality of errors, respectively. Therefore, we extend the lags and incorporate dummy variables to address atypical observations.

The estimation of the revised VAR model is presented in Table \ref{tab:VAR2}. As shown in Table \ref{tab:test-auto_VAR5}, extending the lags from three to five resolves the autocorrelation in errors, and incorporating dummy variables, as shown in Table \ref{tab:week2}, corrects nonnormal errors. The well-specified VAR model for further analysis is:

\begin{equation}
(I_{p} - \pi_{1}L - \pi_{2}L^{2} - \pi_{3}L^{3} - \pi_{4}L^{4} - \pi_{5}L^{5})p_{t} = c + \delta t + \Phi D_{t} + \mu_{t},
\label{eq:VAR1_var5}
\end{equation}

where $D_{t}$ is a vector of five dummy variables: $D_{1}$ takes the value 1 for November 2003 and 0 otherwise; $D_{2}$ takes the value 1 for November 2016 and 0 otherwise; $D_{3}$ takes the value 1 for January 2018 and 0 otherwise; $D_{4}$ takes the value 1 for July 2020 and 0 otherwise; $D_{5}$ takes the value 1 for May 2020 and 0 otherwise. These dummy variables capture atypical data in the model.

\begin{table}[H]
\centering
\caption{Model Estimation VAR(3)}
\begin{tabular}{ccccc} \hline\hline
Variable & log$p^{MW}$ & log$p^{NE}$ & log$p^{SO}$ & log$p^{WE}$ \\ \hline
log$p^{MW}_{t-1}$ & 0.6452 (9.03) & 0.1251 (1.91) & 0.1127 (2.01) & 0.1670 (2.58) \\
log$p^{MW}_{t-2}$ & 0.0603 (0.73) & -0.2625 (-3.49) & -0.1104 (-1.70) & -0.2290 (-3.07) \\
log$p^{MW}_{t-3}$ & 0.0116 (0.15) & 0.0731 (1.07) & -0.1011 (-1.72) & 0.0016 (0.02) \\
log$p^{NE}_{t-1}$ & 0.2057 (2.78) & 0.7392 (10.9) & 0.1941 (3.34) & 0.3858 (5.77) \\
log$p^{NE}_{t-2}$ & -0.2000 (-2.24) & -0.0341 (-0.41) & -0.1852 (-2.65) & -0.2902 (-3.60) \\
log$p^{NE}_{t-3}$ & 0.0243 (0.32) & 0.0964 (1.40) & -0.0018 (-0.03) & -0.0011 (-0.01) \\
log$p^{SO}_{t-1}$ & 0.1127 (1.23) & 0.1528 (1.82) & 0.7743 (10.7) & 0.1040 (1.25) \\
log$p^{SO}_{t-2}$ & -0.0618 (-0.54) & -0.1418 (-1.37) & -0.0921 (-1.03) & -0.1709 (-1.67) \\
log$p^{SO}_{t-3}$ & 0.0277 (0.30) & 0.0976 (1.15) & 0.3104 (4.28) & 0.0025 (0.03) \\
log$p^{WE}_{t-1}$ & 0.1359 (1.78) & 0.1939 (2.79) & 0.1821 (3.04) & 0.6946 (10.0) \\
log$p^{WE}_{t-2}$ & -0.0931 (-1.08) & -0.0322 (-0.41) & -0.0825 (-1.22) & 0.0917 (1.18) \\
log$p^{WE}_{t-3}$ & -0.0165 (-0.23) & -0.0308 (-0.47) & -0.0683 (-1.21) & 0.1082 (1.67) \\
C & 0.2042 (4.06) & 0.0343 (0.74) & 0.0943 (2.39) & 0.1834 (4.02) \\
t & 0.0003 (3.53) & 0.0000 (0.71) & 0.0001 (2.15) & 0.0003 (3.99) \\ \hline\hline
$\Bar{R}^{2}$ & 0.9864 & 0.9908 & 0.9933 & 0.9897 \\
Log-likelihood & 608.03 & 632.64 & 673.92 & 635.03 \\
AIC & -4.36 & -4.54 & -4.85 & -4.56 \\
Schwarz Criterion & -4.18 & -4.36 & -4.66 & -4.38 \\ \hline\hline
Log-likelihood & 2635.35 & & & \\
AIC & -18.96 & & & \\
Schwarz Criterion & -18.22 & & & \\ \hline\hline
\end{tabular}
\label{tab:VAR_original}
\subcaption*{Note: T-statistics are in parentheses.}
\end{table}

\begin{table}[H]
\centering
\caption{Serial Autocorrelation Test, VAR(3)}
\begin{tabular}{ccccccc} \hline \hline
Lag & LRE Stat & df & Prob. & Rao F-Stat & df & Prob. \\ \hline
1 & 25.45 & 16 & 0.06 & 1.60 & (16, 767.5) & 0.06 \\
2 & 27.55 & 16 & 0.03 & 1.73 & (16, 767.5) & 0.03 \\
3 & 31.70 & 16 & 0.01 & 2.00 & (16, 767.5) & 0.01 \\
4 & 18.97 & 16 & 0.27 & 1.18 & (16, 767.5) & 0.27 \\ \hline\hline
\end{tabular}
\subcaption*{Note: The null hypothesis is no serial correlation at lag $h$, where $h = 1, 2, 3, 4$.}
\label{tab:serial_VAR3}
\end{table}

\begin{table}[H]
\caption{Normality Test, Cholesky Covariance}
\begin{subtable}{0.4\textwidth}
\centering
\begin{tabular}{llll} \hline \hline
Comp. & Jarque-Bera & df & Prob \\ \hline
1 & 704.76 & 2 & 0.00 \\
2 & 135.10 & 2 & 0.00 \\
3 & 0.33 & 2 & 0.85 \\
4 & 5.47 & 2 & 0.06 \\
Joint & 1090.25 & 8 & 0.00 \\ \hline \hline
\end{tabular}
\caption{Model VAR(3) with Constants and Trend}
\label{tab:week1}
\end{subtable}
\hfill
\begin{subtable}{0.4\textwidth}
\centering
\begin{tabular}{llll} \hline \hline
Comp. & Jarque-Bera & df & Prob \\ \hline
1 & 4.27 & 2 & 0.11 \\
2 & 1.78 & 2 & 0.41 \\
3 & 2.05 & 2 & 0.35 \\
4 & 6.04 & 2 & 0.05 \\
Joint & 14.15 & 8 & 0.07 \\ \hline \hline
\end{tabular}
\caption{Model VAR(5) with Constant, Trend, and Dummy Variables}
\label{tab:week2}
\end{subtable}
\label{tab:normalidad}
\end{table}

\begin{table}[H]
\centering
\caption{Model VAR(5), Corrected for Normality and Autocorrelation}
\resizebox{.8\textwidth}{!}{
\begin{tabular}{ccccc} \hline\hline
Variable & log$p^{MW}$ & log$p^{NE}$ & log$p^{SO}$ & log$p^{WE}$ \\ \hline
log$p^{MW}_{t-1}$ & 0.6108 (9.23) & 0.1412 (2.50) & 0.1252 (2.35) & 0.1392 (2.22) \\
log$p^{MW}_{t-2}$ & 0.1450 (1.80) & -0.1061 (-1.54) & -0.0012 (-0.02) & -0.1046 (-1.37) \\
log$p^{MW}_{t-3}$ & -0.0827 (-1.03) & 0.0097 (0.14) & -0.1606 (-2.50) & -0.0571 (-0.75) \\
log$p^{MW}_{t-4}$ & 0.1357 (1.72) & -0.0569 (-0.84) & -0.0089 (-0.14) & -0.0068 (-0.09) \\
log$p^{MW}_{t-5}$ & -0.0168 (-0.24) & -0.0088 (-0.14) & -0.0124 (-0.22) & 0.0141 (0.21) \\
log$p^{NE}_{t-1}$ & 0.1395 (2.03) & 0.6659 (11.3) & 0.1151 (2.08) & 0.3113 (4.77) \\
log$p^{NE}_{t-2}$ & -0.0796 (-0.94) & 0.0931 (1.29) & -0.0703 (-1.03) & -0.1882 (-2.35) \\
log$p^{NE}_{t-3}$ & -0.0130 (-0.15) & 0.1132 (1.56) & -0.0380 (-0.55) & -0.1001 (-1.24) \\
log$p^{NE}_{t-4}$ & -0.0057 (-0.06) & -0.0880 (-1.23) & 0.0320 (0.47) & 0.1688 (2.12) \\
log$p^{NE}_{t-5}$ & -0.0112 (-0.16) & 0.0510 (0.87) & -0.0889 (-1.62) & -0.0732 (-1.13) \\
log$p^{SO}_{t-1}$ & 0.1760 (1.97) & 0.1490 (1.95) & 0.7486 (10.4) & 0.1822 (2.14) \\
log$p^{SO}_{t-2}$ & -0.0852 (-0.80) & -0.1724 (-1.90) & -0.0519 (-0.60) & -0.1090 (-1.08) \\
log$p^{SO}_{t-3}$ & 0.0342 (0.32) & 0.0887 (0.99) & 0.1685 (1.99) & -0.0420 (-0.42) \\
log$p^{SO}_{t-4}$ & -0.1573 (-1.50) & 0.0869 (0.97) & 0.1052 (1.25) & 0.0609 (0.61) \\
log$p^{SO}_{t-5}$ & 0.0894 (1.01) & -0.0847 (-1.12) & 0.0296 (0.41) & -0.2001 (-2.38) \\
log$p^{WE}_{t-1}$ & 0.1325 (1.86) & 0.1885 (3.09) & 0.1944 (3.39) & 0.7032 (10.4) \\
log$p^{WE}_{t-2}$ & -0.1040 (-1.25) & -0.1189 (-1.68) & -0.1130 (-1.70) & 0.0182 (0.23) \\
log$p^{WE}_{t-3}$ & -0.0301 (-0.36) & 0.0096 (0.13) & -0.1484 (-2.25) & -0.0442 (-0.56) \\
log$p^{WE}_{t-4}$ & 0.0565 (0.71) & 0.0189 (0.27) & 0.0385 (0.60) & 0.0946 (1.26) \\
log$p^{WE}_{t-5}$ & -0.0415 (-0.62) & 0.0057 (0.10) & 0.1018 (1.91) & 0.0999 (1.59) \\
$D_{1}$ & 0.1503 (3.06) & 0.0207 (0.49) & 0.0486 (1.23) & 0.1772 (3.81) \\
$D_{2}$ & 0.0002 (2.56) & 0.0000 (0.55) & 0.0000 (1.18) & 0.0003 (4.04) \\
$D_{3}$ & 0.1878 (7.66) & 0.1260 (6.02) & 0.0805 (4.08) & 0.0509 (2.19) \\
$D_{4}$ & 0.0364 (1.44) & 0.1125 (5.22) & 0.0714 (3.52) & 0.0686 (2.87) \\
$D_{5}$ & -0.0145 (-0.59) & -0.0716 (-3.44) & 0.0051 (0.26) & -0.0355 (-1.54) \\
C & -0.0332 (-1.35) & 0.1146 (5.47) & 0.0060 (0.30) & -0.0081 (-0.34) \\
t & -0.0823 (-2.89) & -0.0987 (-4.06) & -0.0912 (-3.98) & -0.1139 (-4.22) \\ \hline\hline
$\Bar{R}^{2}$ & 0.98 & 0.99 & 0.99 & 0.98 \\
Log-likelihood & 639.25 & 685.84 & 699.01 & 646.24 \\
AIC & -4.55 & -4.89 & -4.99 & -4.60 \\
Schwarz Criterion & -4.26 & -4.60 & -4.70 & -4.31 \\ \hline\hline
Log-likelihood & 2724.18 & & & \\
AIC & -19.45 & & & \\
Schwarz Criterion & -18.28 & & & \\ \hline\hline
\end{tabular}
}
\label{tab:VAR2}
\subcaption*{Note: T-statistics are in parentheses.}
\end{table}

\begin{table}[H]
\centering
\caption{Serial Autocorrelation Test, VAR(5)}
\begin{tabular}{ccccccc} \hline \hline
Lag & LRE Stat & df & Prob. & Rao F-Stat & df & Prob. \\ \hline
1 & 25.22 & 16 & 0.06 & 1.58 & (16, 721.6) & 0.06 \\
2 & 13.03 & 16 & 0.67 & 0.81 & (16, 721.6) & 0.67 \\
3 & 15.67 & 16 & 0.47 & 0.98 & (16, 721.6) & 0.47 \\
4 & 16.46 & 16 & 0.42 & 1.03 & (16, 721.6) & 0.42 \\ \hline\hline
\end{tabular}
\subcaption*{Note: The null hypothesis is no serial correlation at lag $h$, where $h = 1, 2, 3, 4$.}
\label{tab:test-auto_VAR5}
\end{table}

\begin{table}[H]
\centering
\caption{Stability VAR(5)}
\begin{tabular}{lr} \hline\hline
\multicolumn{1}{l}{Root} & \multicolumn{1}{c}{Modulus} \\ \hline
\multicolumn{1}{l}{0.958379 + 0.035071i} & \multicolumn{1}{c}{0.959020} \\
\multicolumn{1}{l}{0.958379 - 0.035071i} & \multicolumn{1}{c}{0.959020} \\
\multicolumn{1}{l}{0.933121} & \multicolumn{1}{c}{0.933121} \\
\multicolumn{1}{l}{0.850576} & \multicolumn{1}{c}{0.850576} \\
\multicolumn{1}{l}{0.156387 - 0.680552i} & \multicolumn{1}{c}{0.698289} \\
\multicolumn{1}{l}{0.156387 + 0.680552i} & \multicolumn{1}{c}{0.698289} \\
\multicolumn{1}{l}{0.436908 - 0.508710i} & \multicolumn{1}{c}{0.670578} \\
\multicolumn{1}{l}{0.436908 + 0.508710i} & \multicolumn{1}{c}{0.670578} \\
\multicolumn{1}{l}{-0.507077 + 0.430751i} & \multicolumn{1}{c}{0.665337} \\
\multicolumn{1}{l}{-0.507077 - 0.430751i} & \multicolumn{1}{c}{0.665337} \\
\multicolumn{1}{l}{-0.551393 + 0.197171i} & \multicolumn{1}{c}{0.585586} \\
\multicolumn{1}{l}{-0.551393 - 0.197171i} & \multicolumn{1}{c}{0.585586} \\
\multicolumn{1}{l}{-0.231862 + 0.485269i} & \multicolumn{1}{c}{0.537816} \\
\multicolumn{1}{l}{-0.231862 - 0.485269i} & \multicolumn{1}{c}{0.537816} \\
\multicolumn{1}{l}{-0.524999} & \multicolumn{1}{c}{0.524999} \\
\multicolumn{1}{l}{0.103540 + 0.470644i} & \multicolumn{1}{c}{0.481899} \\
\multicolumn{1}{l}{0.103540 - 0.470644i} & \multicolumn{1}{c}{0.481899} \\
\multicolumn{1}{l}{0.229357 + 0.264940i} & \multicolumn{1}{c}{0.350425} \\
\multicolumn{1}{l}{0.229357 - 0.264940i} & \multicolumn{1}{c}{0.350425} \\
\multicolumn{1}{l}{0.281518} & \multicolumn{1}{c}{0.281518} \\ \hline\hline
\end{tabular}
\label{tab:estabilidad_var5}
\end{table}

\subsection{Cointegration Test}

The Johansen cointegration test considers the following characteristics: the price series exhibit a trend, the VAR model includes a constant, and the cointegration vector includes only a constant \footnote{This specification is based on standard empirical work on the Law of One Price.}. The results of the trace and maximum eigenvalue (MEV) cointegration tests are presented in Tables \ref{tab:trace} and \ref{tab:valor propio}. Both indicate only one cointegration relationship, suggesting a stable long-term relationship among the price series.

\begin{table}[H]
\centering
\caption{Cointegration Test Results: Trace}
\begin{tabular}{lrrrr} \hline\hline
\multicolumn{1}{c}{Hypothesis} & \multicolumn{1}{c}{Eigenvalue} & \multicolumn{1}{c}{Trace Statistic} & \multicolumn{1}{c}{0.05 Critical Value} & \multicolumn{1}{c}{Prob.} \\ \hline
None * & 0.099385 & 50.02568 & 47.85613 & 0.0308 \\
At most 1 & 0.057750 & 21.76292 & 29.79707 & 0.3119 \\
At most 2 & 0.018547 & 5.702093 & 15.49471 & 0.7304 \\
At most 3 & 0.002395 & 0.647488 & 3.841466 & 0.4210 \\ \hline\hline
\end{tabular}
\label{tab:trace}
\end{table}

\begin{table}[H]
\centering
\caption{Cointegration Test Results: Maximum Eigenvalue}
\begin{tabular}{lrrrr} \hline\hline
\multicolumn{1}{c}{Hypothesis} & \multicolumn{1}{c}{Eigenvalue} & \multicolumn{1}{c}{Max-Eigen Statistic} & \multicolumn{1}{c}{0.05 Critical Value} & \multicolumn{1}{c}{Prob.} \\ \hline
None * & 0.099385 & 28.26276 & 27.58434 & 0.0409 \\
At most 1 & 0.057750 & 16.06083 & 21.13162 & 0.2213 \\
At most 2 & 0.018547 & 5.054604 & 14.26460 & 0.7349 \\
At most 3 & 0.002395 & 0.647488 & 3.841466 & 0.4210 \\ \hline\hline
\end{tabular}
\label{tab:valor propio}
\end{table}

\subsection{VECM Model Estimation}

Since the VAR model has an order of five, the VECM model is estimated with an order of four \footnote{This is because the VECM model uses the VAR model in first differences.}. The model, in its simplified form, is:

\begin{equation}
\Delta p_{t} = \alpha \beta' \Bar{p}_{t-1} + \Gamma_{1} \Delta y_{t-1} + \Gamma_{2} \Delta y_{t-2} + \Gamma_{3} \Delta y_{t-3} + \Gamma_{4} \Delta y_{t-4} + \gamma + \Phi D_{t} + \mu_{t},
\label{eq:modelo_vecm10}
\end{equation}

where $p_{t}' = [\text{log}p_{t}^{MW}, \text{log}p_{t}^{NE}, \text{log}p_{t}^{SO}, \text{log}p_{t}^{WE}]$, $\Bar{p}_{t}' = [\text{log}p_{t}^{MW}, \text{log}p_{t}^{NE}, \text{log}p_{t}^{SO}, \text{log}p_{t}^{WE}, 1]$, and $D_{t}$ is the vector of dummy variables used in Equation \ref{eq:VAR1_var5}. The term $\gamma$ captures the deterministic trend, and the $\Gamma_{i}$ matrices are $4 \times 4$. The estimation results are presented in Table \ref{tab:vecm_est}.

\begin{table}[H]
\centering
\caption{VECM Model Estimation}
\resizebox{.8\textwidth}{!}{
\begin{tabular}{ccccc} \hline \hline
Cointegration Eq. & & & & \\ \hline
$\text{log}p^{MW}_{t-1}$ & $\text{log}p^{NE}_{t-1}$ & $\text{log}p^{SO}_{t-1}$ & $\text{log}p^{WE}_{t-1}$ & C \\ \hline
1.00 & 1.9085 (4.25) & -1.1484 (-4.19) & -1.7319 (-5.62) & -0.0477 \\ \hline \hline
Error Correction & $\text{d(log}p^{MW})$ & $\text{d(log}p^{NE})$ & $\text{d(log}p^{SO})$ & $\text{d(log}p^{WE})$ \\ \hline
Coint. Eq. 1 & -0.0293 (-1.26) & -0.0665 (-3.42) & -0.0436 (-2.37) & 0.0482 (2.17) \\
$\text{d(log}p^{MW}_{t-1})$ & -0.3070 (-4.44) & 0.1947 (3.36) & 0.1698 (3.10) & 0.1061 (1.60) \\
$\text{d(log}p^{MW}_{t-2})$ & -0.1326 (-1.74) & 0.0791 (1.24) & 0.1686 (2.80) & 0.0047 (0.06) \\
$\text{d(log}p^{MW}_{t-3})$ & -0.2033 (-2.82) & 0.0857 (1.42) & 0.0077 (0.13) & -0.0486 (-0.70) \\
$\text{d(log}p^{MW}_{t-4})$ & -0.0387 (-0.56) & 0.0204 (0.35) & 0.0033 (0.06) & -0.0433 (-0.65) \\
$\text{d(log}p^{NE}_{t-1})$ & 0.1682 (2.29) & -0.1932 (-3.14) & 0.1898 (3.27) & 0.2162 (3.07) \\
$\text{d(log}p^{NE}_{t-2})$ & 0.0662 (0.88) & -0.0996 (-1.58) & 0.1085 (1.82) & 0.0007 (0.01) \\
$\text{d(log}p^{NE}_{t-3})$ & 0.0483 (0.65) & 0.0147 (0.23) & 0.0665 (1.14) & -0.1046 (-1.48) \\
$\text{d(log}p^{NE}_{t-4})$ & 0.0245 (0.35) & -0.0662 (-1.15) & 0.0953 (1.75) & 0.0616 (0.93) \\
$\text{d(log}p^{SO}_{t-1})$ & 0.1352 (1.39) & 0.0653 (0.80) & -0.2882 (-3.77) & 0.2337 (2.52) \\
$\text{d(log}p^{SO}_{t-2})$ & 0.0483 (0.49) & -0.1092 (-1.33) & -0.3366 (-4.35) & 0.1220 (1.30) \\
$\text{d(log}p^{SO}_{t-3})$ & 0.0703 (0.73) & -0.0139 (-0.17) & -0.1587 (-2.09) & 0.0941 (1.02) \\
$\text{d(log}p^{SO}_{t-4})$ & -0.0983 (-1.10) & 0.0724 (0.97) & -0.0490 (-0.69) & 0.1515 (1.77) \\
$\text{d(log}p^{WE}_{t-1})$ & 0.1070 (1.42) & 0.0857 (1.36) & 0.1340 (2.25) & -0.1578 (-2.18) \\
$\text{d(log}p^{WE}_{t-2})$ & 0.0053 (0.07) & -0.0269 (-0.42) & 0.0221 (0.37) & -0.1240 (-1.72) \\
$\text{d(log}p^{WE}_{t-3})$ & -0.0143 (-0.19) & -0.0170 (-0.27) & -0.1271 (-2.16) & -0.1628 (-2.29) \\
$\text{d(log}p^{WE}_{t-4})$ & 0.0355 (0.53) & -0.0005 (-0.00) & -0.0904 (-1.71) & -0.0803 (-1.25) \\
C & 0.0023 (1.49) & 0.0021 (1.66) & 0.0032 (2.61) & 0.0021 (1.43) \\
$D_{1}$ & 0.1910 (7.71) & 0.1263 (6.08) & 0.0811 (4.14) & 0.0562 (2.37) \\
$D_{2}$ & 0.0434 (1.70) & 0.1080 (5.06) & 0.0689 (3.42) & 0.0622 (2.55) \\
$D_{3}$ & -0.0119 (-0.48) & -0.0741 (-3.58) & 0.0052 (0.26) & -0.0372 (-1.57) \\
$D_{4}$ & -0.0309 (-1.24) & 0.1149 (5.51) & 0.0056 (0.28) & -0.0052 (-0.22) \\
$D_{5}$ & -0.0894 (-3.11) & -0.0977 (-4.06) & -0.0924 (-4.07) & -0.1163 (-4.22) \\ \hline \hline
$\Bar{R}^{2}$ & 0.3387 & 0.4473 & 0.3602 & 0.3459 \\
Log-likelihood & 634.13 & 681.95 & 697.55 & 645.59 \\
AIC & -4.52 & -4.88 & -4.99 & -4.61 \\
Schwarz Criterion & -4.22 & -4.57 & -4.69 & -4.30 \\ \hline \hline
Log-likelihood & 2712.71 & & & \\
AIC & -19.38 & & & \\
Schwarz Criterion & -18.10 & & & \\ \hline\hline
\end{tabular}
}
\label{tab:vecm_est}
\subcaption*{Note: T-statistics are in parentheses.}
\end{table}

In equilibrium, where $\hat{\beta}' \Bar{p}_{t} = 0$, the long-term vector is:

\begin{equation}
\text{log}p_{t}^{MW} = -1.9085 \text{log}p_{t}^{NE} + 1.1484 \text{log}p_{t}^{SO} + 1.7319 \text{log}p_{t}^{WE} + 0.0477,
\end{equation}

where the coefficients, being logarithms of prices, can be interpreted as elasticities. The variables $\text{log}p^{NE}$, $\text{log}p^{SO}$, and $\text{log}p^{WE}$ adjust against an imbalance \footnote{With only one cointegration relationship, a weak exogeneity test is unnecessary, as the t-statistics of each coefficient indicate whether it is statistically different from zero. This does not apply when $r > 1$.}.

\begin{figure}[H]
\centering
\caption{Cointegration Graph}
\includegraphics[scale=0.7]{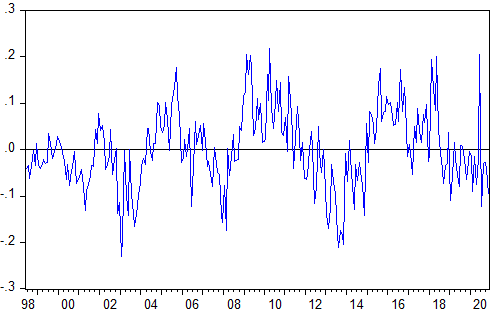}
\label{fig:cointegracion_grap}
\subcaption*{Note: The vertical axis shows the value of the vector, and the horizontal axis represents time.}
\end{figure}

Figure \ref{fig:cointegracion_grap} shows the cointegration graph, which graphically demonstrates stationary behavior.

\subsection{Granger Causality Estimation}

Cointegration analysis establishes a long-term relationship among prices, but it does not imply causality. Granger causality exists when a variable $X$ improves the prediction of $Y$, i.e., the coefficients of the lags of $X$ are statistically different from zero. The joint null hypothesis \footnote{Assuming the coefficients $\hat{\beta}_0 = \hat{\beta}_1 = \dots = \hat{\beta}_k$ correspond to $X$.} that $X$ does not Granger-cause $Y$ ($H_0: \hat{\beta}_0 = \hat{\beta}_1 = \dots = \hat{\beta}_k = 0$) is tested.

Since the series are integrated of order one and exhibit one cointegration relationship, the Granger causality test, presented in Table \ref{tab:granger}, is based on the VECM model estimated in Equation \ref{eq:modelo_vecm10}. The results indicate:
\begin{itemize}
\item No region Granger-causes $\text{log}p^{MW}$.
\item The price logarithm of the Midwest region Granger-causes $\text{log}p^{NE}$.
\item The price logarithms of the Midwest, Northeast, and West regions Granger-cause $\text{log}p^{SO}$.
\item The price logarithm of the Northeast region Granger-causes $\text{log}p^{WE}$.
\end{itemize}

\begin{table}[H]
\centering
\caption{Granger Causality Test Results}
\begin{tabular}{ccccc} \hline\hline
Dep. Variable: D$(\text{log}p^{MW})$ & & & & \\ \hline
Excluded & Chi-sq & df & Prob. & Null Hypothesis \\ \hline
D$(\text{log}p^{NE})$ & 5.33 & 4 & 0.25 & Not rejected \\
D$(\text{log}p^{SO})$ & 4.63 & 4 & 0.32 & Not rejected \\
D$(\text{log}p^{WE})$ & 2.68 & 4 & 0.61 & Not rejected \\
All & 20.35 & 12 & 0.06 & \\ \hline
Dep. Variable: D$(\text{log}p^{NE})$ & & & & \\ \hline
Excluded & Chi-sq & df & Prob. & Null Hypothesis \\ \hline
D$(\text{log}p^{MW})$ & 12.22 & 4 & 0.01 & Rejected \\
D$(\text{log}p^{SO})$ & 5.98 & 4 & 0.20 & Not rejected \\
D$(\text{log}p^{WE})$ & 2.93 & 4 & 0.56 & Not rejected \\
All & 33.03 & 12 & 0.001 & \\ \hline
Dep. Variable: D$(\text{log}p^{SO})$ & & & & \\ \hline
Excluded & Chi-sq & df & Prob. & Null Hypothesis \\ \hline
D$(\text{log}p^{MW})$ & 14.73 & 4 & 0.005 & Rejected \\
D$(\text{log}p^{NE})$ & 13.65 & 4 & 0.008 & Rejected \\
D$(\text{log}p^{WE})$ & 14.23 & 4 & 0.006 & Rejected \\
All & 62.22 & 12 & 0.00 & \\ \hline
Dep. Variable: D$(\text{log}p^{WE})$ & & & & \\ \hline
Excluded & Chi-sq & df & Prob. & Null Hypothesis \\ \hline
D$(\text{log}p^{MW})$ & 4.27 & 4 & 0.36 & Not rejected \\
D$(\text{log}p^{NE})$ & 16.99 & 4 & 0.001 & Rejected \\
D$(\text{log}p^{SO})$ & 7.94 & 4 & 0.09 & Not rejected \\
All & 52.05 & 12 & 0.00 & \\ \hline\hline
\end{tabular}
\label{tab:granger}
\end{table}

\subsection{Law of One Price}

Since only one cointegration vector is identified, the Law of One Price (LOP) does not hold simultaneously across all four regions, as $N-1$ cointegration relationships (where $N$ is the number of series) are required for simultaneous compliance. Thus, the hypothesis test in Equation \ref{eq:ho_lop} cannot be applied. However, the results allow testing the LOP for pairwise price comparisons, i.e., $p_{it} = p_{jt}$, by imposing restrictions on the cointegration vector in the bivariate case.

The likelihood ratio test results in Table \ref{tab:test_lop} show that the LOP holds only for the Northeast and West regions at a significance level of $\alpha = 0.01$.

\begin{table}[H]
\centering
\caption{Testing the Law of One Price Through Pairwise Comparisons}
\begin{tabular}{cccc} \hline\hline
Region $i$ & Region $j$ & Hypothesis & LR Test (with $r=1$) \\ \hline
Midwest & Northeast & $\beta' = [1, -1, 0, 0, *]$ & 18.38 (0.0003) \\
Midwest & South & $\beta' = [1, 0, -1, 0, *]$ & 19.73 (0.0001) \\
Midwest & West & $\beta' = [1, 0, 0, -1, *]$ & 16.75 (0.0007) \\
Northeast & South & $\beta' = [0, 1, -1, 0, *]$ & 22.44 (0.0000) \\
Northeast & West & $\beta' = [0, 1, 0, -1, *]$ & 0.04 (0.0450) \\
South & West & $\beta' = [0, 0, 1, -1, *]$ & 19.74 (0.0001) \\ \hline\hline
\end{tabular}
\label{tab:test_lop}
\subcaption*{Note: * refers to the constant of the cointegration vector, which is unconstrained. P-values are in parentheses.}
\end{table}

\section{Comments and Discussions}

This paper examines the Law of One Price (LOP) in U.S. regional beef markets and investigates causal interdependence among regional prices using cointegration and Granger causality analysis.

Our findings indicate that only one cointegration vector exists among the four regional retail beef price series. This suggests that regional prices do not fully converge in the long run, implying that the LOP does not hold simultaneously across all U.S. regions. This partial integration may reflect frictions such as transportation costs, differences in retail pricing practices, infrastructure disparities, or limited arbitrage opportunities.

However, pairwise testing reveals that the Northeast and West regions share a price relationship consistent with the LOP, suggesting stronger market integration between these regions. This may be due to similarities in demand patterns, distribution networks, or supply chain structures. The presence of integration between only certain regions aligns with previous literature \cite{Rumankova}, which highlights partial LOP compliance in segmented markets.

Granger causality analysis provides further insights into the direction of price influences (see Table \ref{tab:granger}). Notably, the South region is significantly influenced by price movements in the Midwest, Northeast, and West, despite having the highest concentration of beef cattle (44.6\% as of 2019  \cite{ERS2019}). This suggests that high production levels do not necessarily confer price leadership, possibly due to structural constraints such as centralized retail pricing or greater exposure to national supply chains.

Overall, the results indicate that while U.S. regional beef markets are interrelated, full price integration is absent. The observed asymmetries and partial convergence suggest that regional characteristics and institutional factors shape price dynamics. One potential factor is the lack of coordinated national pricing or infrastructure policies, which may hinder market unification.

This study has limitations. It focuses solely on retail prices for beef fillet, a premium, low-volume cut, which may exhibit higher volatility and be less representative of broader market behavior. Additionally, the use of BLS price data, which reflect listed rather than transaction-based or volume-weighted prices, may introduce measurement errors. The analysis also excludes transportation costs, quality differences, and wholesale-retail margins, which could influence regional price dynamics.

Future research should address these limitations by incorporating other beef cuts (e.g., ground beef), transaction-level data at higher frequencies (e.g., scanner data), and explicit modeling of cost structures and spatial frictions using spatial econometric models. Such extensions would enable a more comprehensive evaluation of market integration and the extent to which the LOP holds under realistic conditions.

\end{document}